\documentclass[12pt]{article}
\usepackage{epsfig}
\begin{document}
\large

\par
\noindent {\bf Higgs Mechanism in the Standard Model and a
Possibility of its Direct Physical Realization}
\par
\begin{center}
\vspace{0.3cm} Beshtoev Kh. M. (beshtoev@cv.jinr.ru)
\par
\vspace{0.3cm} Joint Institute for Nuclear Research, Joliot Curie
6, 141980 Dubna, Moscow region, Russia.
\end{center}
\vspace{0.3cm}

\par
Abstract \\

\par
The aim of this work was to answer the question: Is the direct
physical realization of the Higgs mechanism possible? It is shown
that this mechanism cannot have a direct physical realization
since the condition for this realization is not fulfilled. It
means that if in the new collider at CERN a scalar particle is
detected, it does not mean that it is a Higgs particle. \\

\par
\section{Introduction}

At present the generally accepted point of view is that the
standard electroweak model [1] has full confirmation and now
remains only to detect a Higgs scalar particle.
\par
The lagrangian of the standard electroweak model [1] besides the
quark and lepton interactions via $W, Z$ bosons, also includes
Higgs sector which is used to generate lepton, quark and $W, Z$
boson masses. This mechanism provides the renormalizability of
this model [2]. At present three families of quarks and leptons as
well as $W, Z$ bosons have been detected and their masses (with
the exception of neutrinos) have been measured [3]. It is
necessary to remark that the attempt to register the scalar Higgs
boson was not successful. In connection with the commission of the
new collider at CERN, where this scalar particle can be
registered, the problem of origin of elementary particle masses
becomes important. The strong and electromagnetic interactions can
generate the masses of elementary particles since they are
left-right symmetrical. In contrast to these interactions the weak
interactions are left-side interactions but not the left-right
symmetrical ones. As a result of this the Higgs mechanism is used
to generate masses in the standard electroweak model.
\par
In principle masses of elementary particles can be generated in an
interaction in analogy with the chromodynamics. This approach was
applied in the Technicolor model [4,5].
\par
At present it is a very important question whether the scalar
particle (if discovered) is a Higgs particle or it has another
origin.
\par
This work is devoted to the discussion of a possible direct
physical realization of the Higgs mechanism where the Higgs scalar
particle appears.

\par
\noindent
\section{Higgs Mechanism in the Standard Model and
a Possibility of its Direct Physical Realization}

\par
\noindent \subsection{Higgs Mechanism in the Standard Model}

\par
A doublet of scalar Higgs fields
$$
\Phi  = \left( \begin{array}{c} \Phi ^{(+)}\\ \Phi
^{(o)}\end{array}\right) , \eqno(1)
$$
with hypercharge equal to the unity, is introduced. It is assumed
that this doublet interacts with the vector and fermion fields in
such a way that local gauge invariance is not broken. To the
Lagrangian of the electroweak interactions we add the Higgs
potential $V(\Phi ^{+}, \Phi)$
$$
V(\Phi ^{+}, \Phi ) = k(\Phi^{+}, \Phi )^{2} - \mu ^{2}(\Phi^{+},
\Phi ) , \eqno(2)
$$
($k, \mu^{2}$ are positive constants), which leads to vacuum
degeneracy and to a non vanishing vacuum expectation value
$<\Phi^{o}>$ of the field $\Phi ^{o}$:
$$
<\Phi ^{o}> = \sqrt{{\mu ^{2}\over 2k}} = \frac{\nu}{\sqrt{2}}
,\qquad \nu = \sqrt{\frac{\mu^2}{k}}, \eqno(3)
$$
this means that (fixing the vacuum state) we can generate a mass
term of the fields of the intermediate bosons, fermions, and a
Higgs boson.
\par
 In the unitary gauge by using (3) we can rewrite $V(\Phi)$ in
the following form ($\nu^2 = \frac{\mu^2}{k}$):
 $$
 V(\Phi) = - \frac{\mu^4}{2} (\nu + \Phi^o)^2 + \frac{k}{4} (\nu + \Phi^o)^4
$$
$$
= -\frac{\mu^4}{4 k} + \mu^2 (\Phi^o)^2 + ... = -\frac{\mu^4}{4 k}
+ \frac{m_\Phi^2}{2} (\Phi^o)^2  + ...    , \eqno(4)
$$
hence, we see that Higgs boson $\Phi^o$ has mass $m_{\Phi^o}^2 =
2 \mu^2$.
\par
The covariant derivative for Higgs fields is
$$
D_\alpha \Phi = (\partial_\alpha - i g \frac{\tau^i A^i_\alpha}{2}
- i \frac{g'}{2} B_\alpha) \Phi . \eqno(5)
$$
The kinetic energy term of Higgs bosons (in the unitary gauge) has
the following form:
$$
(D^\alpha \Phi)^{+} D_\alpha \Phi = M^2_W W^{\alpha {+}}
W^{-}_\alpha + \frac{M^2_Z}{2} Z^\alpha Z_\alpha + ... , \eqno(6)
$$
where $W^{\pm}_\alpha = (A^1_\alpha \pm A^2_\alpha)/ \sqrt{2}$,
and their masses are as follows:
$$
M^2_W = g^2 \frac{\nu^2}{4},\quad M^2_Z = (g^2 + g'^2)
\frac{\nu^2}{4} .
$$
\par
The quark masses are obtained by using a Lagrangian of the Yukawa
type which is $SU(2)_ {L} \times U(1)$ invariant:
$$
{\cal L}_{1} = - \sum^{3}_{i; q=d,s,b} \bar \Psi_{iL} M^{1}_{iq}
q_{R} \bar \Phi + H. C.  , \eqno(7)
$$
$$
{\cal L}_{2} = -  \sum^{3}_{i; q=u,c,t} \bar \Psi_{iL} M^{2}_{iq}
q_{R} \bar \Phi + H. C. ,
$$
where $M^{1}, M^{2}$ - complex $3 \times 3$ matrix, and $\bar
\Phi$
$$
\bar \Phi = i\tau_{2} \Phi ^{*} = \left(\begin{array}{c} \Phi^{o*} \\
-\Phi^{+*} \end{array} \right) , \eqno(8)
$$
is a doublet of Higgs fields with hypercharge $Y = -1$.
\par
Taking into account (3) and using the gauge invariance of the
Lagrangian (4), (8), we can choose (in the unitary gauge)
$$
\Phi (x) = \left(\begin{array}{c} 0\\ {{\nu  + \Phi^{o}(x)}\over
\sqrt{2}}
\end{array} \right) ,\qquad
\bar\Phi (x) = \left(\begin{array}{c}{{\nu  + \Phi^{o}(x)}\over
\sqrt{2}}\\0
\end{array}\right) ,
\eqno(9)
$$
where $\Phi^{o}(x)$ is the neutral scalar Higgs field.
\par
Substituting (9) in (7) for the quark masses we obtain the
expressions
$$
{\cal L}_{1} = - \bar p_{L} {M'}_{1} p_{R} + H. C. , \eqno(10)
$$
$$
{\cal L}_{2} = - \bar n_{L} {M'}_{2} n_{R}  + H. C. ,
$$
where
$$
p_{L,R} = \left(\begin{array}{c} u_{L,R}\\ c_{L,R}\\
t_{L,R}\end{array}\right) ,\qquad n_{L,R} = \left(\begin{array}{c}
d_{L,R}\\ s_{L,R}\\ b_{L,R}
\end{array}\right)  .
$$
Thus, the elements ${M'}_{1}, {M'}_{2}$ of the quark mass matrix
are equal to the constants of the quark-Higgs-boson Yukawa
coupling up to the factor $ \nu $.\\

\par
\noindent
\subsection{Remarks to the Higgs Mechanism in the
Electroweak Model and a Possibility of its Direct Physical
Realization}

\par
We know that quarks, leptons and vector bosons have the same
masses in every point of the Universe. Then Higgs fields must fill
the Universe  and since the masses are real masses, then the Higgs
fields must also be real (here we have the analogy with the
superconductivity). If the Higgs field is real, then the energy
density of this field  is $\rho_{Higgs} \sim 2\cdot 10^{49}
GeV/cm^{3}$ [6, 7] (see also references in [7]). It is a huge
value. The measured energy density in the Universe is $\rho_{Univ}
\sim  10^{-4} GeV/cm^{3}$. Then the relation of the energy density
of the Higgs fields to the measured energy value is
$$
\rho_{Higgs}/ \rho_{Univ} \sim 10^{53} . \eqno(11)
$$
It is interesting to remark that at this density of the energy the
condition to create a black hole is fulfilled for the volume with
radius
$$
R \geq \sqrt{\frac{3}{4 \pi \rho_{Higgs} G_N}} \approx 9.5 \quad
cm , \eqno(12)
$$
where $G_N$ is a gravitational constant (i.e., the Universe will
be filled with the black holes). It is clear that the Higgs
mechanism has big problems for its realization.
\par
The other problem: Is there a possibility of its direct physical
realization? The Higgs potential which is added to the lagrangian
of electroweak interactions is given by expression (2). The
lagrangian of the Higgs field is
$$
L(\Phi) = \frac{1}{2} \partial_{\mu} \Phi^{+} \partial^{\mu} \Phi
+ V(\Phi^{+}, \Phi) . \eqno(13)
$$
Before going on our discussion let us consider mechanisms of
generation of particle effective masses.
\par
1. Usually it is supposed that, when we ascribe a charge to a
particle, its masse changes. We cannot compute this changing but
by using the renormalization group we can compute the changing of
this mass dependence of the momenta transfer.
\par
2. When the confined (bound) states of particles are formed then
the particle effective masses change. This changing of the
particle effective masses has a local character.
\par
3. There are other mechanisms of the effective masses changing,
for example: the change of the effective mass of electrons in
metal [8], or correlation of electrons in the superconductive
state [9] (see also-Wikipedia: Superconductivity). In these cases
the changing of the electron effective mass (or correlation of
electrons) has a non local character and it changes in all this
medium.
\par
The main problem is: Which type of the mass generation does the
Higgs mechanism belong to?
\par
It is clear that quarks, leptons and gauge bosons must have the
same masses in every point of the Universe. If their masses are
generated as the result of their interactions with Higgs field,
then in order to get masses, the sources of the Higgs field must
be distributed in all the Universe uniformly. And in every point
of the Unverse without the Higgs field the particle masses will be
equal to zero (i.e., they will have zero masses).
\par
The problem is: How and in what way does the Higgs field fill the
Unverse? This problem is just a physical problem. Then the sources
of the Higgs field must be distributed continuously or in the form
of lattices. Without further discussion it is obvious that these
distributions of the Higgs field in the Universe are not
realistic. Especially if to take into account what enormous vacuum
energy density appears in this mechanism, we have to come to a
conclusion that this mechanism is physically inadmissible in spite
of the fact that this mechanism provides the electroweak model to
be renormalized.
\par
As we have seen above the Higgs mechanism cannot be realized
physically and then the nature of this scalar particle will remain
unclear. It is necessary to remark that if in the new collider at
CERN the scalar particle is detected it will not mean that it is
just a Higgs particle. Besides, as it is stressed in [10]
 that the Higgs mechanism contains a contradiction. There are some arguments
that mass sources can be a mechanism which is analogous to the
strong interactions [11], i.e., masses are generated via
interactions between the quark and lepton subparticles, then the
problem of singularity of the theory does not arise (i.e., it will
be solved in analogy with chromodynamics). \\

\section{Conclusion}

It is well known that the Higgs mechanism in the electroweak model
is perfect from the mathematical point of view and and it leads to
renormalizability of this model. It allows one to make
computations of higher orders of the perturbation theory.
\par
The aim of this work was to show: Is the direct physical
realization of this mechanism possible. It is shown that this
mechanism cannot have direct physical realization since the
condition for this realization is not fulfilled. It means that if
in the new collider at CERN a scalar particle is detected it does
not mean that it is just a Higgs particle.
\par
The central problem of the weak interactions is:
\par
1. Why are these interactions left-right non-symmetric (i.e., why
at interac\-tions via $W$ bosons the right components of fermions
do not participate). The electroweak model is only a model to
compute weak and electromagnetic processes but it does not give
the answer to the question above. The next question is:
\par
2. What is a dynamical source of the particle (i.e., quarks and
leptons) masses? No doubt, that finding (discovering) schemes to
compute of weak processes was a very important problem. But also a
very important problem is to understand the basis of the weak
interactions.
\par
It is necessary to stress that there is an another problem. It is
well known that in the theories with left-right symmetric
interactions we can use renormalization groups for computation of
the couple constants in dependence of square momenta transfer.
\par
3. Is it correct using the renormalization group in the case when
the theory is not left-right symmetrical as it take place in the
weak interactions [12]?
\\

{\bf References}

\begin{bibliography}{999}

\par
\noindent
1. Glashow S.L.  Nucl. Phys. 1961, vol.22, p.579 ;
\par
Weinberg S., Phys.  Rev. Lett., 1967, vol.19, p.1264 ;
\par
Salam A., Proc. of the 8-th    Nobel  Symp.,  edited  by     N.
Svarthholm
\par
(Almgvist and Wiksell,  Stockholm) 1968,p.367.
\par
\noindent 2. t'Hooft G., Nucl. Physics, 1971,, v. B33, p. 173;
1971,, v. B35, p. 167;
\par
Review of part. Prop., J. of Physics G (Nuclear and Particle
Physics), 2006, v.33, p.110.
\par
\noindent 3. Review of part. Prop., J. of Physics G (Nuclear and
Particle Physics), 2006, v.33, p.31.
\par
\noindent 4. Weinberg S., Phys. Rev. 1976, v.D19, p.974; 1979,
v.D19, p.1271;
\par
Sussind L., Phys. Rev. 1979, v.D20, p.2619;
\par
Eichen E. and Lane K., Phys. Lett., 1980, v.B90, p.125.
\par
\noindent 5. Dimopulos S. and Susskind L., Nuclear Physics, 1980,
v. B155, p.237;
\par
Dimopulos S., Nuclear Physics, 1980, v. B168, p.69.
\par
\noindent 6. Higgs P.W., Phys. Lett., 1964, Vol.12, p.132;
Phys.Rev., 1966,
\par
vol.145, p.1156;
\par
Englert F., Brout R.- Phys. Rev. Lett, 1964,
\par
vol.13,p.321;
\par
Guralnik G.S., Hagen C.R.,  Kible  T.W.B.-  Phys.  Rew.  Lett,
1964,
\par
vol.13,p.585.
\par
\noindent 7. Langacker p., Phys. Rep. 1981, v.72, N4.
\par
\noindent 8. J. Zaiman, Principles of Solid State Physics,  M.,
Nauka, 1974 y.;
\par
Ch. Kittel, Introduction to Solid State Physics, M., Nauka, 1978
y.
\par
\noindent 9. J. Shriffer , Theory of  Superconductivity , M.,
Nauka, 1970 y.;
\par
M. Tinkham, Introduction to Superconductivity, M., Nauka, 1980 y.;
\par
N. M. Plakida, High-Temperature Superconductivity, Springer,
\par
1995 y.
\par
\noindent 10. Beshtoev Kh. M. and Toth L., JINR Commun. E2-90-475,
\par
Dubna, 1990;
\par
Beshtoev Kh. M., JINR Commun. E2-92-195, Dubna,1992.
\par
\noindent 11. H. Harari Phys. Let., 1979, v. 86B, p. 83;
\par
M. A. Shupe, Phys. Let., 1979, v. 86B, p. 87;
\par
Beshtoev Kh. M., JINR Commun. P2-83-735, Dubna,1983;
\par
Beshtoev Kh. M., JINR Commun. E2-99-137, Dubna,1999.
\par
\noindent 12. Beshtoev Kh. M., hep-ph/0703252v.1, March, 2007.

\end{bibliography}

\end{document}